\documentclass[prl,twocolumn,showpacs,superscriptaddress,floatfix]{revtex4}
\usepackage{graphicx}
\usepackage{amssymb}
\usepackage{citesort}

\begin{document}
\title{Intermediate phase of the one dimensional half-filled Hubbard-Holstein
model}
\author{R.T. Clay and R.P. Hardikar}
\affiliation{Department of Physics and Astronomy and ERC Center for 
Computational Sciences, Mississippi State University, Mississippi State MS 39762}
\date{\today}
\begin{abstract}
We present a detailed numerical study of the Hubbard-Holstein model in
one dimension at half filling, including full finite-frequency quantum
phonons. At half filling, the effects of the electron-phonon and
electron-electron interactions compete, with the Holstein phonon
coupling acting as an effective negative Hubbard onsite interaction
$U$ that promotes on-site electron pairs and a Peierls charge-density
wave state.  Most previous work on this model has assumed that only
Peierls or $U>0$ Mott insulator phases are possible at half filling. However, there
has been speculation that a third metallic phase exists between the
Peierls and Mott phases.  We present results confirming the
intermediate metallic phase, and show that the Luttinger liquid
correlation exponent $K_\rho>1$ in this region, indicating dominant
superconducting pair correlations.  We explore the full phase diagram
as a function of onsite Hubbard $U$, phonon coupling constant, and
phonon frequency.
\end{abstract}

\pacs{71.10.Fd, 71.30.+h, 71.45.Lr}\maketitle

Electron-phonon (e-ph) interactions can give rise to a number of
interesting effects in low-dimensional materials, including
superconductivity as well as charge-density wave and insulating
phenomena.  Frequently these materials feature strong
electron-electron (e-e) interactions as well, leading to very rich
phase diagrams that combine lattice, charge, and spin (magnetic)
orderings. We will focus specifically on materials where the electrons
are coupled to localized vibrational modes, which may be of relatively
high frequency.  This type of e-ph interaction is most studied in
molecular crystal materials, including the quasi-one- and
quasi-two-dimensional organic superconductors\cite{Ishiguro} and
fullerene superconductors \cite{Gunnarsson97a}.  In all of these
materials, a fundamental question is whether the effects of e-e and
e-ph interactions compete or cooperate with each other. In this
Letter, we examine this issue within one of the most basic models. We
find that despite the two interactions each separately favoring
insulating states, {\it together} they can mediate an unexpected
metallic phase with superconducting (SC) pair correlations.

The model we consider is the one-dimensional (1D) Hubbard-Holstein
model (HHM), with Hamiltonian
\begin{eqnarray}
H&=& -t \sum_{j,\sigma} (c^\dagger_{j+1,\sigma}c_{j,\sigma} + h.c.) + U\sum_j 
n_{j,\uparrow} n_{j,\downarrow} \label{ham} \\
&+& g \sum_{j,\sigma} (b_j^\dagger + b_j) n_{j,\sigma} + \omega \sum_j b^\dagger_j b_j, \nonumber
\end{eqnarray}
where $c^\dagger_{j,\sigma}$ ($c_{j,\sigma}$) are fermionic creation
(annihilation) operators for electrons on site $j$ with spin $\sigma$,
$b^\dagger_{j}$ ($b_{j}$) are bosonic creation (annihilation)
operators for phonons at site $j$, and
$n_{j,\sigma}=c^\dagger_{j,\sigma} c_{j,\sigma}$. The dispersionless
phonons have frequency $\omega$ and are coupled to the local electron
density with coupling strength $g$ \cite{Holstein59a}. $U$ is the
Hubbard on-site e-e interaction energy. All energies will be given
below in units of the hopping integral $t$.

The properties of the 1D 1/2-filled HHM are well understood in two
limits: the static or $\omega\rightarrow 0$ and the $\omega\rightarrow
\infty$ limit \cite{Hirsch83a}. First, in the static limit, the ground
state is Peierls distorted for any nonzero e-ph coupling $g=0^+$. As
the phonons couple to the electron density, the Peierls state is a
2k$_F$ charge-density wave (CDW) consisting of alternating large and
small site charges.  In the $\omega\rightarrow\infty$ limit, the
retarded interaction between electrons mediated by the phonons becomes
instantaneous in imaginary time, and the phonons may be integrated
out.  This leads to an effective renormalized Hubbard interaction
$U_{\rm{eff}}=U-2g^2/\omega$. While strictly at
$\omega\rightarrow\infty$ the Peierls state cannot occur, for finite
$\omega$ the Peierls state may again occur, although the mapping to an
effective negative $U$ is no longer exact. However, based on the
$\omega\rightarrow\infty$ mapping, it has been believed that the
ground state phase of the HHM may be determined through $U_{\rm{eff}}$.
$U_{\rm{eff}}$ should
correspond to the $U>0$ Hubbard model, which in 1D at 1/2 filling has
a finite charge gap for any $U>0$ and no spin gap. We shall refer to
this state as the Mott state. On the other hand, if $U_{\rm{eff}}<0$,
e-ph interactions dominate over e-e interactions, and the ground state
is Peierls CDW distorted.

Hence at $\omega=0$, the Peierls distortion occurs unconditionally,
while at $\omega=\infty$ there is no Peierls state.  Much less is
known in the intermediate $\omega$ region. In the $U=0$ model, the
Peierls state may be viewed as a traditional band insulator for small
$\omega$, and as bipolaronic insulator composed of tightly bound pairs
in the large $\omega$ limit, with a crossover between these pictures
for $\omega\sim t$ \cite{Fehske04a}.  If $\omega>0$, the Peierls
distortion only occurs for e-ph coupling $g$ above some critical value
$g_c$ \cite{Wu95a,Jeckelmann99a}.  For $g<g_c$, the ground state is
then metallic at 1/2 filling (at $U=0$).  For $U=0$, $g_c$ has been
calculated using a functional integral method \cite{Wu95a} and also
via the density matrix renormalization group (DMRG)
\cite{Jeckelmann99a}. For intermediate $U$ as well as $\omega$, far
less is known.  However, it has been recently proposed that a {\it
metallic} ground state exists {\it intermediate} between the Mott and
Peierls states at 1/2 filling, i.e. when $U_{\rm{eff}}$ is close to
zero\cite{Takada03a}.  This metallic phase occurs for intermediate
$\omega$ and hence cannot be predicted from the small and large
$\omega$ limits. Our goal in this Letter is to confirm
 this metallic state and investigate its properties.  
Its existence is perhaps not surprising given the known
existence of a metallic phase for $g<g_c$ at $U=0$; this region of the
phase diagram continues to exist for finite $U$. We further show that
this metallic region persists for a substantial range of parameters
provided $\omega$ is not too small.
\begin{figure}[tb]
\resizebox{3.2in}{!}{\includegraphics{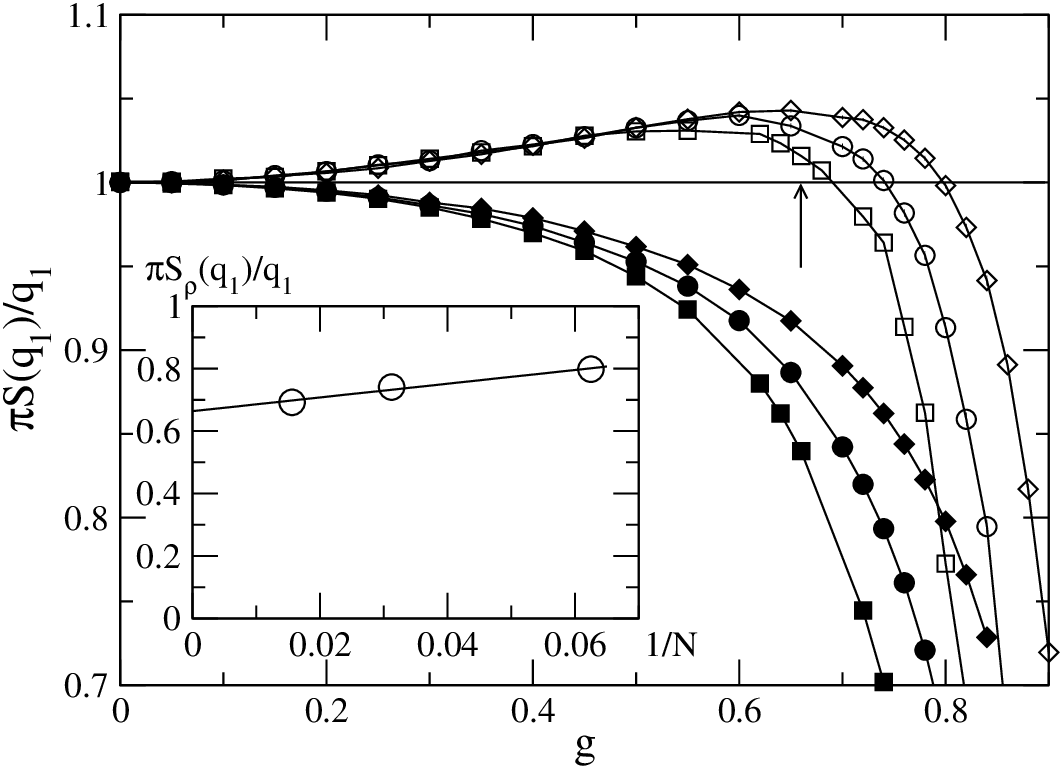}}
\caption{$U=0$, $\omega=1$ results for long-wavelength charge (open
symbols) and spin (filled symbols) structure factors versus $g$ for
periodic systems of $N=16$ (diamonds), $N=32$ (circles), and $N=64$
(squares) sites. Statistical errors are smaller than the symbols. The
inset shows finite-size scaling of the critical coupling $g_c$
(indicated by arrow) where $K_\rho=1$.}
\label{figu0}
\end{figure}

Hamiltonian Eq.~\ref{ham} is difficult to analyze in the intermediate
coupling region due to the presence of both electrons and phonons. The
numerical method we use is the Stochastic Series Expansion (SSE)
quantum Monte Carlo method with directed loops
\cite{Syljuasen02a}. SSE is a powerful method for non-frustrated
quantum spin systems or 1D electron lattice models where no sign
problem occurs. Importantly, there are no approximations in the method
besides finite system size and temperature.  Electron-phonon
interactions have been incorporated in SSE for both spin
models\cite{Sandvik99b} and electron models\cite{Sengupta03a}.  As in
these references, we treat the phonons in the occupation-number basis.
The number of phonons per lattice site is unbounded in the
thermodynamic limit, but for a finite system at a finite temperature,
the number of phonons may be truncated. We choose this truncation in a
similar manner as the truncation of sequence length in the SSE
method\cite{Syljuasen02a}: in the equilibration phase of the
calculation, the phonon truncation is increased to exceed the current
number of phonons on any given lattice site whenever necessary.  All
results shown below used periodic lattices of $N$ sites, with inverse
temperatures of at least $\beta/t=2N$ and phonon cutoffs of 
up to 30 phonons per lattice
site.  In this Letter we focus on results for the HHM model, and
details on the SSE implementation will be published separately. Our
code was checked extensively against Lancz\"{o}s exact diagonalization
results for several different observables.  We also implemented the
quantum parallel tempering algorithm \cite{Sengupta02a}, where
different processors of a parallel computer are assigned different
model parameters ($U$, $g$, and $\omega$).  A Metropolis probability
is then computed to switch configurations between processors with
adjacent parameters. As in reference \onlinecite{Sengupta02a}, we
found this technique essential in obtaining smooth data across quantum
phase transition boundaries.  

The low energy properties of any 1D gapless interacting electron model 
may be mapped to an effective continuum model, or
Luttinger Liquid (LL).  The properties of the LL, and in particular
the decay with distance of different correlation functions are then
described by two correlation exponents, $K_\rho$ for charge
properties, and $K_\sigma$ for spin \cite{Voit95a}.  $K_\rho$ values
greater than 1 indicate dominant attractive SC correlations (no SC
long-range order is possible in strictly 1D systems), while $K_\rho<1$
corresponds to repulsive charge correlations.  For models with
spin-rotation symmetry, $K_\sigma$ is always equal to 1.  $K_\sigma=0$
then indicates the presence of a spin gap.  These exponents are most
easily computed via the SSE data from the static structure factors:
\begin{equation}
S_{\rho,\sigma}(q)=\frac{1}{N}\sum_{j,k} e^{iq(j-k)} \langle 
(n_{j\uparrow}
\pm n_{j\downarrow}) (n_{k\uparrow}\pm n_{k\downarrow}) \rangle
\end{equation}
$K_\rho$ and $K_\sigma$  are then proportional to the
slope of the corresponding structure factor in
the long wavelength limit $q\rightarrow 0$ \cite{Clay99a}:
\begin{equation}
K_{\rho,\sigma}=\frac{1}{\pi q}S_{\rho,\sigma}(q\rightarrow 0)
\end{equation}
These values once finite-size scaled may then be used to determine the
quantum phase boundaries.  A second observable we will use are the
charge and spin stiffnesses, $\rho_c$ and $\rho_s$, measured in the
SSE method via the winding number \cite{Sengupta02a}. A zero stiffness
indicates a gap in the corresponding sector, while nonzero $\rho$
indicates no gap. We also verified directly that charge-charge
(spin-spin) correlation functions showed staggered order in the Peierls
(Mott) phases.

We first present results for $U=0$ and $\omega=1$. Fig.~\ref{figu0}
shows the slope of charge and spin structure factors $\pi
S_{\rho,\sigma}(q_1)/q_1$ evaluated at the smallest wavevector
$q_1=2\pi/N$, plotted versus e-ph coupling $g$.  For $g\rightarrow 0$
both $K_\rho$ and $K_\sigma$ tend to exactly 1, as required for free
electrons.  Finite size effects become very small in this limit,
consistent with the expected vanishing of logarithmic corrections when
$K_\rho=K_\sigma=1$ \cite{Sengupta03a,Eggert96a}.  At a critical
coupling $g_c$, $K_\rho$ crosses 1 and tends to zero, marking the
transition to the Peierls state. From finite-size scaling of 16, 32,
and 64 site systems vs. $1/N$ (see Fig.~\ref{figu0} inset), we find
the critical coupling $g_c=0.66\pm 0.01$, with the uncertainty 
estimated from the linear fit. A previous DMRG
study, which did not attempt finite-size scaling found $g_c\approx
0.8$ \cite{Jeckelmann99a}.  For $g<g_c$, the LL exponent $K_\rho>1$,
and $K_\sigma$ scales to zero with increasing $N$. Therefore, for
$U=0$ and $g<g_c$, the ground state is metallic, with dominant SC pair
correlations and a spin gap.

\begin{figure}[tb]
\resizebox{3.2in}{!}{\includegraphics{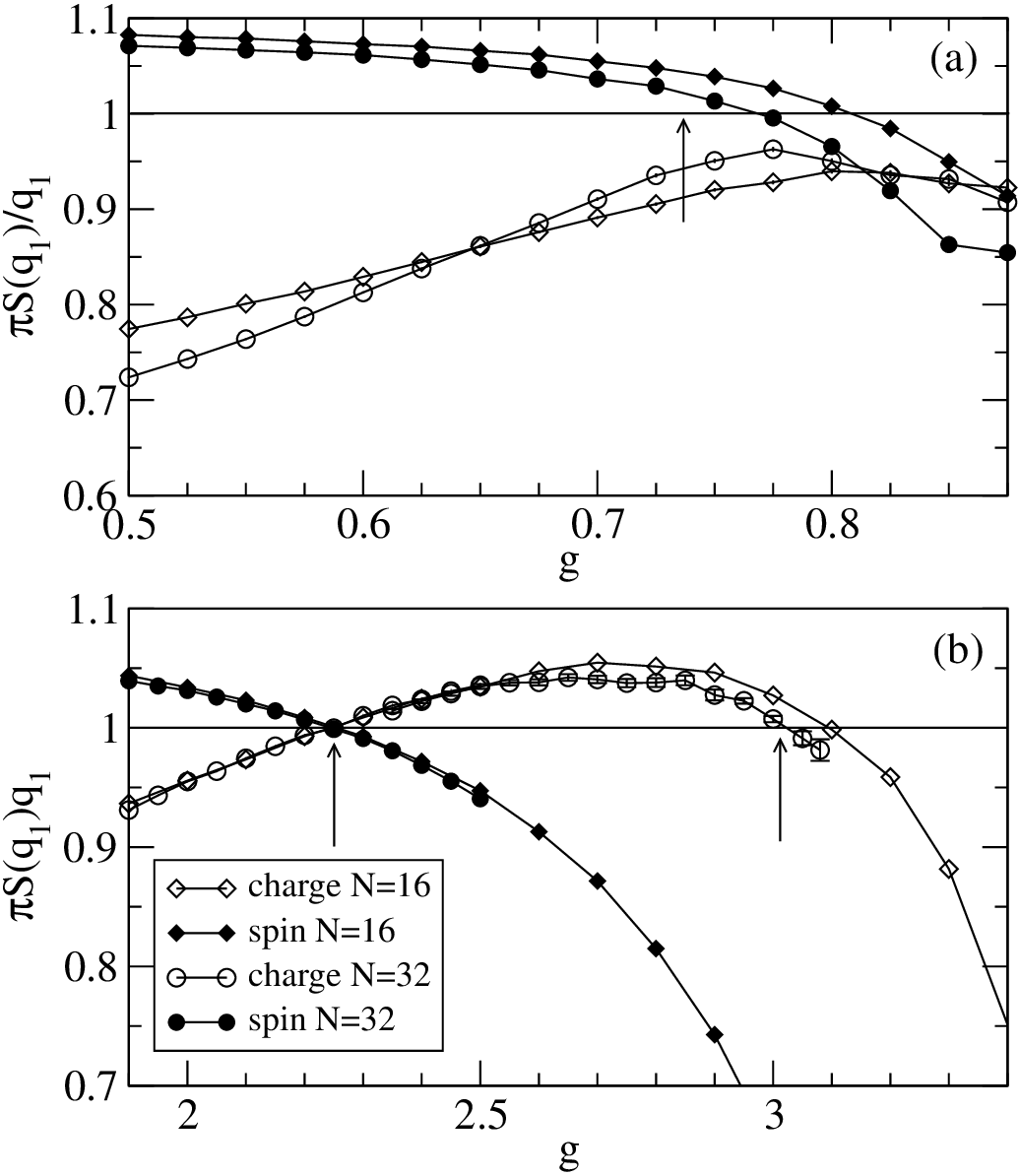}}
\caption{Long-wavelength charge (open symbols) and spin (filled
symbols) structure factors versus $g$. (a) $U=2$, $\omega=$0.5; arrow
indicates finite-size scaled $g_c$ (b) $U$=2, $\omega$=5; arrows
indicate finite-size scaled transitions $g_{c1}$ and $g_{c2}$.}
\label{figu2}
\end{figure}
For any finite $U$ at $g=0$ and 1/2 filling, the dominant ground state
correlations of Eq.~\ref{ham} are spin-density wave (SDW). In this
ground state $K_\sigma=1$, and $K_\rho=0$ indicating an insulating
state with no spin gap. As this 
\begin{figure}[tb]
\resizebox{3.2in}{!}{\includegraphics{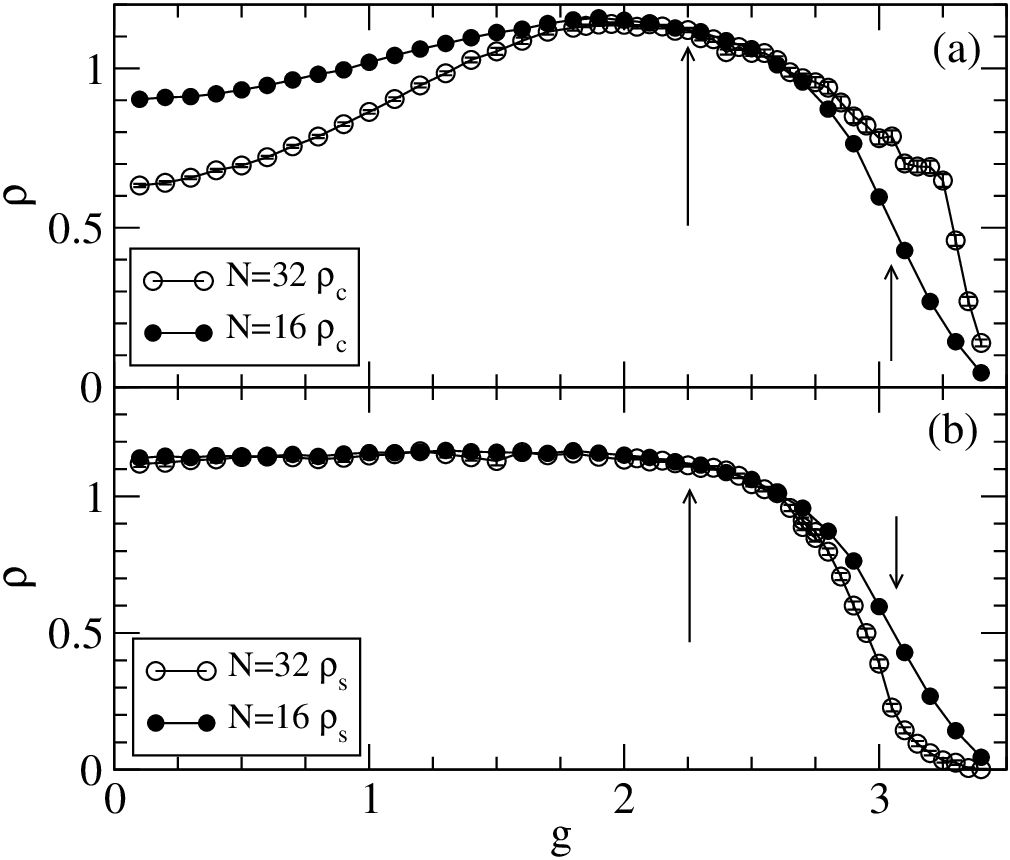}}
\caption{Charge (a) and spin (b) stiffnesses versus $g$ for $U$=2, $\omega$=5.
Open (closed) symbols are for $N$=32 (16) site lattices.
Arrows mark the transition points $g_{c1}$ and $g_{c2}$ determined
from Fig.~\ref{figu2}(b).}
\label{figstiff}
\end{figure}
Mott state is {\it not} equivalent to
the metallic state found for $g<g_c$ in Fig.~\ref{figu0}, we may then
expect the possibility of {\it three} different phases for the HHM
with $U>0$ and $g>0$: Mott, Peierls, and metallic SC. Fig.~\ref{figu2}
shows the charge and spin structure factors again versus $g$ with
$U=2$.  We first discuss results for small phonon frequency
$\omega=0.5$ shown in Fig.~\ref{figu2}(a).  We find that for small
$g$, $K_\sigma$ tends toward 1 as system size increases. At a critical
coupling $g_c$, $K_\sigma$ becomes less than 1, indicating the opening
of a spin gap and the transition to the Peierls state. The charge
exponent $K_\rho$ scales towards zero on both sides of the transition,
developing a peak at $g=g_c$.  Similar results are found for the
1/2-filled 1D extended Hubbard model (EHM), where $K_\rho$ also peaks
at the transition between CDW and bond-order wave (BOW) phases
\cite{Sandvik04a}.  The finite value of $K_\rho$ on the boundary
indicates that the transition is of continuous nature
\cite{Sandvik04a}. Like the CDW-(BOW)-SDW transition in the 1/2-filled
EHM, we find similar behavior for small $\omega$ in the HHM as $U$
increases: the value of $K_\rho$ at the transition {\it decreases}
with $U$, indicating that for large $U$ the transition is first order
rather than continuous.  The correspondence is not surprising, as the
effect of a nearest-neighbor interaction $V\gg U$ can be viewed as an
effective negative $U$\cite{Clay99a}. 

Charge and spin response typical for large $\omega$ are shown in
Fig.~\ref{figu2}(b), here shown for $U=2$ and $\omega=5$.  Again, for
small $g$, we find $K_\sigma$ tends toward 1 as system size increases,
and $K_\rho$ tends toward 0, consistent with the Mott state. However,
at a critical coupling $g_{c1}$ $K_\rho$ exceeds 1 and $K_\sigma$
becomes less than one, indicating the dominant-SC state found for
$U=0$, $g<g_c$.  The value of $g_{c1}$ is very close to
the expected value where $U_{\rm{eff}}=0$. Increasing
$g$ further, for $g=g_{c2}$ $K_\rho$ becomes less than 1, indicating
the opening of a charge gap and the Peierls state.  In the
intermediate region $g_{c1}<g<g_{c2}$, the properties of the model are
identical to the $U=0$ model for $g<g_c$ as seen in Fig.~\ref{figu0},
i.e. metallic with dominant SC correlations. While the second
transition point $g_{c2}$ is finite-size dependent, finite-size
effects are very weak at the first transition since
$K_\rho=K_\sigma=1$.  It is then clear that due to the {\it crossing}
of $K_\rho$ and $K_\sigma$ curves at exactly 1 at $g=g_{c1}$, a region
of $K_\rho>1$ {\it must} exist for $g>g_{c1}$.

\begin{figure}[tb]
\resizebox{3.0in}{!}{\includegraphics{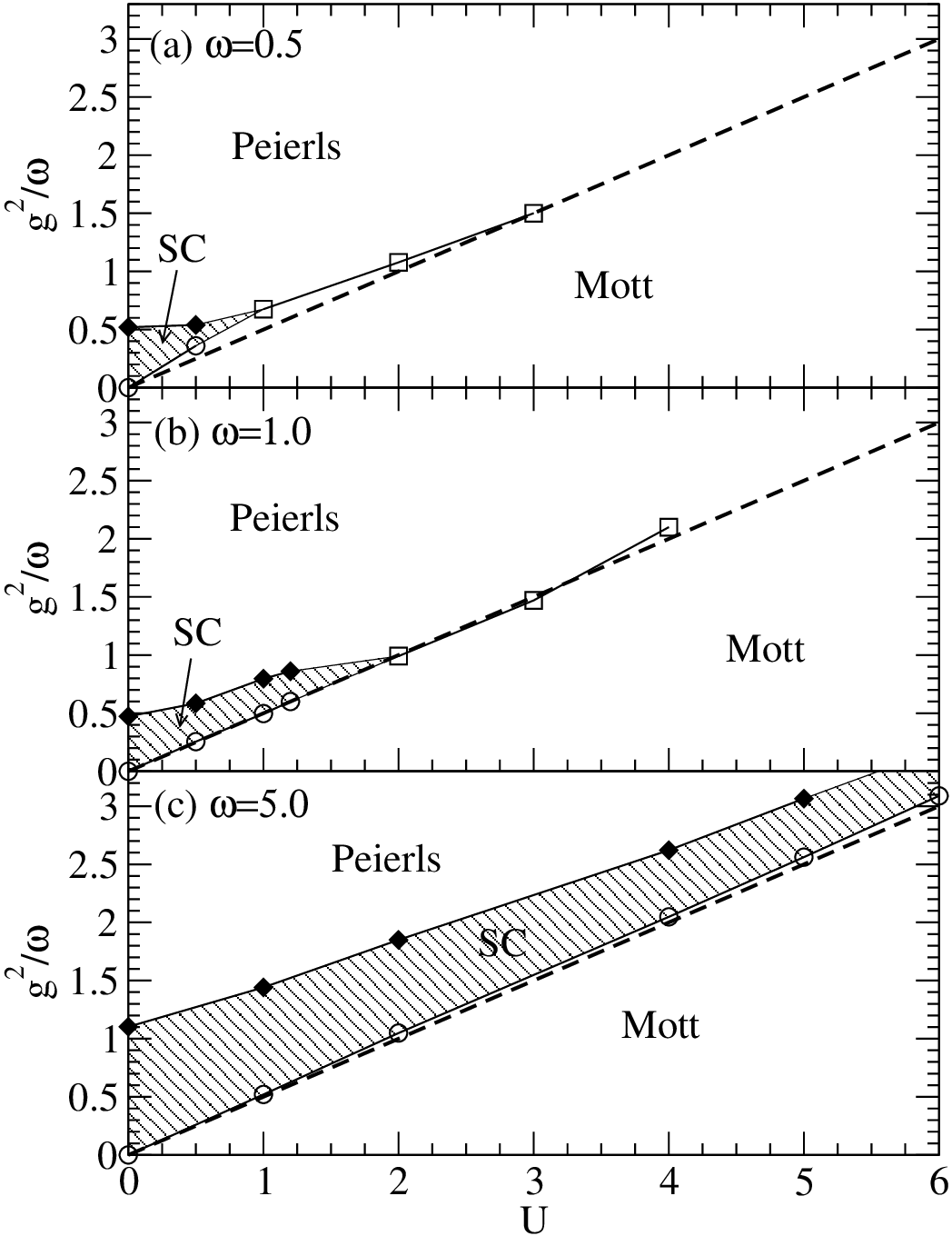}}
\caption{Evolution of the intermediate phase versus $\omega$.
(a) $\omega$=0.5 (b) $\omega$=1 (c)
$\omega$=5. Open circles indicate the boundary $g_{c1}$ between Mott
and SC phases, filled diamonds the boundary $g_{c2}$
between SC and Peierls phases, and open squares the boundary
between Mott and Peierls phases. Dashed line 
indicates $U=2g^2/\omega$. }
\label{figpd}
\end{figure}
To confirm the structure factor results indicating two transitions at
$g_{c1}$ and $g_{c2}$, we present the charge ($\rho_c$) and spin
($\rho_s$) stiffnesses in Fig.~\ref{figstiff}. As with the structure
factors, the presence of two transitions may only be detected from
$\rho_c$ and $\rho_s$ after finite-size scaling has been
performed. For clarity, we plot only two system sizes in
Fig.~\ref{figstiff}.  In Fig.~\ref{figstiff}(a) we see that for small
$g$, $\rho_c$ decreases with system size, indicating a charge
gap. $\rho_s$ is nearly constant with system size, indicating no spin
gap. In the intermediate phase, this is reversed, with $\rho_c$
remaining constant or increasing with system size, and $\rho_s$
decreasing with system size. This again indicates a spin gap but no
charge gap in the SC region. Finally, for $g>g_{c2}$, both stiffnesses
go to zero in the Peierls state.

Fig.~\ref{figpd} shows the evolution of the intermediate phase as a
function of $\omega$, with phase boundaries determined from
finite-size scaling of 16, 24, and 32 site structure factor data.  In
all cases, the SC phase exists for $g<g_c$ exactly at $U=0$. As $U$ is
increased at fixed $\omega$, the SC region then shrinks. For small
enough $\omega$, we see two different sequences of phases, either
Mott-SC-Peierls for small $U$ (as in Fig.~\ref{figu2}(b)), or
Mott-Peierls (as in Fig.~\ref{figu2}(a)) for large $U$.  Numerically
it is difficult to precisely determine point where all three phases
meet, but we find that the SC phase disappears at $U\approx 1$ for
$\omega=0.5$, and at $U\approx 2$ for $\omega=1.0$.  In
Fig.~\ref{figpd}(a) and (b), we have not plotted points for the
Mott/Peierls boundary for large $U$, as in this region the transition
becomes strongly first order, making exact determination of the
boundary difficult. However, the transition appears to remain close to
$g^2/\omega=U/2$.  For $\omega=5$, we were not able to 
access large enough $U$ to determine the
upper cutoff $U$ necessary to suppress the SC state, but the
intermediate phase appears to persist up to at least $U\approx 7$
for $\omega=5$.  In all cases, we
find the line $U_{\rm{eff}}=0$ to very accurately predict either the
Mott-SC or Mott-Peierls boundary, but not the SC-Peierls boundary.
Our phase diagram is slightly different from
reference \onlinecite{Takada03a}, where the metallic region was found
to extend to either side of the $U_{\rm{eff}}=0$ line.

In conclusion, we have shown that a metallic SC region exists
intermediate between Peierls and Mott phases in the 1D 1/2-filled HHM.
While in the 1/2-filled model with small phonon frequency $\omega<t$
the intermediate phase occupies a relatively small region of the phase
diagram, we expect that in the presence of doping (in fact most of the
organic SC's are 1/4-filled \cite{Ishiguro}), the size of this region
will be greatly enhanced. Furthermore, while the SC pairing found here
exactly at 1/2-filling consists of on-site pairs or bipolarons, with
doping both onsite and nearest-neighbor pairing may be mediated by the
Holstein phonons\cite{Bonca00a}.  We are currently exploring these
possibilities in the doped model.

This work was supported by Oak Ridge Associated Universities.  We
thank A.~Sandvik and P.~Sengupta for discussions regarding the
SSE method.  Numerical calculations were performed at
the Mississippi State University ERC Center for Computational Sciences.


\begin{thebibliography}{17}
\expandafter\ifx\csname natexlab\endcsname\relax\def\natexlab#1{#1}\fi
\expandafter\ifx\csname bibnamefont\endcsname\relax
  \def\bibnamefont#1{#1}\fi
\expandafter\ifx\csname bibfnamefont\endcsname\relax
  \def\bibfnamefont#1{#1}\fi
\expandafter\ifx\csname citenamefont\endcsname\relax
  \def\citenamefont#1{#1}\fi
\expandafter\ifx\csname url\endcsname\relax
  \def\url#1{\texttt{#1}}\fi
\expandafter\ifx\csname urlprefix\endcsname\relax\def\urlprefix{URL }\fi
\providecommand{\bibinfo}[2]{#2}
\providecommand{\eprint}[2][]{\url{#2}}

\bibitem[{\citenamefont{Ishiguro et~al.}(1998)\citenamefont{Ishiguro, Yamaji,
  and Saito}}]{Ishiguro}
\bibinfo{author}{\bibfnamefont{T.}~\bibnamefont{Ishiguro}},
  \bibinfo{author}{\bibfnamefont{K.}~\bibnamefont{Yamaji}}, \bibnamefont{and}
  \bibinfo{author}{\bibfnamefont{G.}~\bibnamefont{Saito}},
  \emph{\bibinfo{title}{Organic Superconductors}}
  (\bibinfo{publisher}{Springer-Verlag}, \bibinfo{address}{New York},
  \bibinfo{year}{1998}).

\bibitem[{\citenamefont{Gunnarsson}(1997)}]{Gunnarsson97a}
\bibinfo{author}{\bibfnamefont{O.}~\bibnamefont{Gunnarsson}},
  \bibinfo{journal}{Rev.\ Mod.\ Phys.} \textbf{\bibinfo{volume}{69}},
  \bibinfo{pages}{575} (\bibinfo{year}{1997}).

\bibitem[{\citenamefont{Holstein}(1959)}]{Holstein59a}
\bibinfo{author}{\bibfnamefont{T.}~\bibnamefont{Holstein}},
  \bibinfo{journal}{Ann. Phys.} \textbf{\bibinfo{volume}{8}},
  \bibinfo{pages}{325} (\bibinfo{year}{1959}).

\bibitem[{\citenamefont{Hirsch and Fradkin}(1983)}]{Hirsch83a}
\bibinfo{author}{\bibfnamefont{J.~E.} \bibnamefont{Hirsch}} \bibnamefont{and}
  \bibinfo{author}{\bibfnamefont{E.}~\bibnamefont{Fradkin}},
  \bibinfo{journal}{Phys.\ Rev.\ B} \textbf{\bibinfo{volume}{27}},
  \bibinfo{pages}{1680} (\bibinfo{year}{1983}).

\bibitem[{\citenamefont{Fehske et~al.}(2004)\citenamefont{Fehske, Wellein,
  Hager, Wei\protect{$\beta$}e, and Bishop}}]{Fehske04a}
\bibinfo{author}{\bibfnamefont{H.}~\bibnamefont{Fehske}},
  \bibinfo{author}{\bibfnamefont{G.}~\bibnamefont{Wellein}},
  \bibinfo{author}{\bibfnamefont{G.}~\bibnamefont{Hager}},
  \bibinfo{author}{\bibfnamefont{A.}~\bibnamefont{Wei\protect{$\beta$}e}},
  \bibnamefont{and} \bibinfo{author}{\bibfnamefont{A.~R.}
  \bibnamefont{Bishop}}, \bibinfo{journal}{Phys.\ Rev.\ B}
  \textbf{\bibinfo{volume}{69}}, \bibinfo{pages}{165115}
  (\bibinfo{year}{2004}).

\bibitem[{\citenamefont{Wu et~al.}(1995)\citenamefont{Wu, Huang, and
  Sun}}]{Wu95a}
\bibinfo{author}{\bibfnamefont{C.}~\bibnamefont{Wu}},
  \bibinfo{author}{\bibfnamefont{Q.}~\bibnamefont{Huang}}, \bibnamefont{and}
  \bibinfo{author}{\bibfnamefont{X.}~\bibnamefont{Sun}},
  \bibinfo{journal}{Phys.\ Rev.\ B} \textbf{\bibinfo{volume}{52}},
  \bibinfo{pages}{R15683} (\bibinfo{year}{1995}).

\bibitem[{\citenamefont{Jeckelmann et~al.}(1999)\citenamefont{Jeckelmann,
  Zhang, and White}}]{Jeckelmann99a}
\bibinfo{author}{\bibfnamefont{E.}~\bibnamefont{Jeckelmann}},
  \bibinfo{author}{\bibfnamefont{C.}~\bibnamefont{Zhang}}, \bibnamefont{and}
  \bibinfo{author}{\bibfnamefont{S.}~\bibnamefont{White}},
  \bibinfo{journal}{Phys.\ Rev.\ B} \textbf{\bibinfo{volume}{60}},
  \bibinfo{pages}{7950} (\bibinfo{year}{1999}).

\bibitem[{\citenamefont{Takada and Chatterjee}(2003)}]{Takada03a}
\bibinfo{author}{\bibfnamefont{Y.}~\bibnamefont{Takada}} \bibnamefont{and}
  \bibinfo{author}{\bibfnamefont{A.}~\bibnamefont{Chatterjee}},
  \bibinfo{journal}{Phys.\ Rev.\ B} \textbf{\bibinfo{volume}{67}},
  \bibinfo{pages}{081102R} (\bibinfo{year}{2003}).

\bibitem[{\citenamefont{Syljuasen and Sandvik}(2002)}]{Syljuasen02a}
\bibinfo{author}{\bibfnamefont{O.~F.} \bibnamefont{Syljuasen}}
  \bibnamefont{and} \bibinfo{author}{\bibfnamefont{A.~W.}
  \bibnamefont{Sandvik}}, \bibinfo{journal}{Phys.\ Rev.\ E}
  \textbf{\bibinfo{volume}{66}}, \bibinfo{pages}{046701}
  (\bibinfo{year}{2002}).

\bibitem[{\citenamefont{Sandvik and Campbell}(1999)}]{Sandvik99b}
\bibinfo{author}{\bibfnamefont{A.~W.} \bibnamefont{Sandvik}} \bibnamefont{and}
  \bibinfo{author}{\bibfnamefont{D.~K.} \bibnamefont{Campbell}},
  \bibinfo{journal}{Phys.\ Rev.\ Lett.} \textbf{\bibinfo{volume}{83}},
  \bibinfo{pages}{195} (\bibinfo{year}{1999}).

\bibitem[{\citenamefont{Sengupta et~al.}(2003)\citenamefont{Sengupta, Sandvik,
  and Campbell}}]{Sengupta03a}
\bibinfo{author}{\bibfnamefont{P.}~\bibnamefont{Sengupta}},
  \bibinfo{author}{\bibfnamefont{A.~W.} \bibnamefont{Sandvik}},
  \bibnamefont{and} \bibinfo{author}{\bibfnamefont{D.~K.}
  \bibnamefont{Campbell}}, \bibinfo{journal}{Phys.\ Rev.\ B}
  \textbf{\bibinfo{volume}{67}}, \bibinfo{pages}{245103}
  (\bibinfo{year}{2003}).

\bibitem[{\citenamefont{Sengupta et~al.}(2002)\citenamefont{Sengupta, Sandvik,
  and Campbell}}]{Sengupta02a}
\bibinfo{author}{\bibfnamefont{P.}~\bibnamefont{Sengupta}},
  \bibinfo{author}{\bibfnamefont{A.~W.} \bibnamefont{Sandvik}},
  \bibnamefont{and} \bibinfo{author}{\bibfnamefont{D.~K.}
  \bibnamefont{Campbell}}, \bibinfo{journal}{Phys.\ Rev.\ B}
  \textbf{\bibinfo{volume}{65}}, \bibinfo{pages}{155113}
  (\bibinfo{year}{2002}).

\bibitem[{\citenamefont{Voit}(1995)}]{Voit95a}
\bibinfo{author}{\bibfnamefont{J.}~\bibnamefont{Voit}}, \bibinfo{journal}{Rep.
  Prog. Phys.} \textbf{\bibinfo{volume}{58}}, \bibinfo{pages}{977}
  (\bibinfo{year}{1995}).

\bibitem[{\citenamefont{Clay et~al.}(1999)\citenamefont{Clay, Sandvik, and
  Campbell}}]{Clay99a}
\bibinfo{author}{\bibfnamefont{R.~T.} \bibnamefont{Clay}},
  \bibinfo{author}{\bibfnamefont{A.~W.} \bibnamefont{Sandvik}},
  \bibnamefont{and} \bibinfo{author}{\bibfnamefont{D.~K.}
  \bibnamefont{Campbell}}, \bibinfo{journal}{Phys.\ Rev.\ B}
  \textbf{\bibinfo{volume}{59}}, \bibinfo{pages}{4665} (\bibinfo{year}{1999}).

\bibitem[{\citenamefont{Eggert}(1996)}]{Eggert96a}
\bibinfo{author}{\bibfnamefont{S.}~\bibnamefont{Eggert}},
  \bibinfo{journal}{Phys.\ Rev.\ B} \textbf{\bibinfo{volume}{54}},
  \bibinfo{pages}{R9612} (\bibinfo{year}{1996}).

\bibitem[{\citenamefont{Sandvik et~al.}(2004)\citenamefont{Sandvik, Balents,
  and Campbell}}]{Sandvik04a}
\bibinfo{author}{\bibfnamefont{A.~W.} \bibnamefont{Sandvik}},
  \bibinfo{author}{\bibfnamefont{L.}~\bibnamefont{Balents}}, \bibnamefont{and}
  \bibinfo{author}{\bibfnamefont{D.~K.} \bibnamefont{Campbell}},
  \bibinfo{journal}{Phys.\ Rev.\ Lett.} \textbf{\bibinfo{volume}{92}},
  \bibinfo{pages}{236401} (\bibinfo{year}{2004}).

\bibitem[{\citenamefont{\protect{Bon\u{c}a}
  et~al.}(2000)\citenamefont{\protect{Bon\u{c}a}, \protect{Katra\u{s}nik}, and
  Trugman}}]{Bonca00a}
\bibinfo{author}{\bibfnamefont{J.}~\bibnamefont{\protect{Bon\u{c}a}}},
  \bibinfo{author}{\bibfnamefont{T.}~\bibnamefont{\protect{Katra\u{s}nik}}},
  \bibnamefont{and} \bibinfo{author}{\bibfnamefont{S.~A.}
  \bibnamefont{Trugman}}, \bibinfo{journal}{Phys.\ Rev.\ Lett.}
  \textbf{\bibinfo{volume}{84}}, \bibinfo{pages}{3153} (\bibinfo{year}{2000}).
\end{thebibliography}
\end{document}